# Online Political Microtargeting: Promises and Threats for Democracy

Frederik J. Zuiderveen Borgesius, Judith Möller, Sanne Kruikemeier, Ronan Ó Fathaigh, Kristina Irion, Tom Dobber, Balazs Bodo, Claes de Vreese*

## 1. Introduction

Political campaigns are increasingly combining data-driven voter research with personalised political advertising: online political microtargeting.[1] Through political microtargeting, a political party can identify the individual voters which it is most likely to convince. Additionally, a party can match its message to the specific interests and vulnerabilities of these voters. Modern online marketing techniques promise to make microtargeting even more tailored to individual voters, and more effective. These techniques are primarily used in the United States, but have recently gained popularity in European countries too. Several parties in European countries are looking into the possibilities of microtargeting.

Online political microtargeting is a type of personalised communication that involves collecting information about people, and using that information to show them targeted political advertisements. Politicians apply microtargeting because they expect that targeting makes ads more effective. Such ads can address issues which are important to an individual, adapting the format and language to meet the individual needs and interests for maximum effect. Recipients of targeted political information are more likely to act upon it. Online political microtargeting may be both a blessing and a curse to democracies. It could increase participation, and lead to more knowledge among voters about certain topics. But microtargeting also brings risks. For instance, a political party could, misleadingly, present itself as a one-issue party to different individuals. And data collection for microtargeting raises privacy concerns.

This paper focuses on the following questions: what is online political microtargeting, and what are its promises and threats? This paper combines insights from both a legal and social science perspective. We focus mostly on European countries and the US. Section 2 introduces the practice of online political microtargeting. Section 3 discusses the promises of online political microtargeting, and Section 4 the threats. Section 5 discusses why the threats, while serious, should not be overstated. Section 6 explores how policymakers in European countries could intervene, and sketches some problems they would encounter if they wanted to intervene. Section 7 concludes: we call for more research and debate about online political microtargeting.

---

\* Dr. F.J. Zuiderveen Borgesius is a researcher at the Vrije Universiteit Brussels, LSTS Research Group on Law Science Technology & Society. Dr. B. Bodo, R. Ó Fathaigh, and Dr. K. Irion are affiliated to the University of Amsterdam, IViR Institute for Information Law. Dr. J. Möller, T. Dobber Dr. S. Kruikemeier, and Prof. C. de Vreese are affiliated to the University of Amsterdam, ASCoR Amsterdam School of Communication Research. Corresponding author: F.J. Zuiderveen Borgesius, email: fzuiderv@vub.ac.be.

1 We thank the editors of the Utrecht Law Review, the anonymous peer reviewers, Prof. Natali Helberger, and the other members of the Personalised Communications team for their helpful suggestions. This work was supported by the European Research Council under Grant 638514 (PersoNews).

## 2. Online political microtargeting

In this paper, we focus on *online* political microtargeting, a category of political microtargeting. Online political microtargeting involves 'creating finely honed messages targeted at narrow categories of voters' based on data analysis 'garnered from individuals' demographic characteristics and consumer and lifestyle habits'.[2] Online political microtargeting can take the 'form of political direct marketing in which political actors target personalized messages to individual voters by applying predictive modelling techniques to massive troves of voter data'.[3] Online political microtargeting could also be seen as a type of behavioural advertising, namely political behavioural advertising. Behavioural advertising is a modern marketing technique that involves tracking people's online behaviour to use the collected information to display individually targeted advertisements.[4]

Online political microtargeting is used, for example, to identify voters who are likely to vote for a specific party and therefore can be targeted with mobilising messages. (For ease of reading, we also refer to 'microtargeting'). Microtargeting also enables a political party to select policy stances that match the interests of the targeted voter – for instance family aid for families, or student benefits for students.

### 2.1 Online political microtargeting in the US

In the US, political microtargeting has developed in the context of offline canvassing.[5] Yet, there has been a rise in data-driven campaigning and a sophistication of microtargeting that was unimaginable just a few decades ago. Political scientist Colin Bennett highlights four trends that can help to explain the rise in political microtargeting in the US: 'the move from voter management databases to integrated voter management platforms; the shift from mass-messaging to micro-targeting employing personal data from commercial data brokerage firms; the analysis of social media and the social graph; and the decentralization of data to local campaigns through mobile applications.'[6]

In the US, political parties and intermediaries hold extremely detailed information about possible voters.[7] New methods of voter data collection and data analysis have improved and enriched traditional forms of political microtargeting like canvassing. But these possibilities have also enabled much more refined methods of *online* political marketing, which is the focus of this paper.

In the US, several companies offer online microtargeting services especially to politicians. For instance, companies like CampaignGrid and Cambridge Analytica enable politicians to target people with ads on Facebook, LinkedIn, and elsewhere on the web.[8] Cambridge Analytica claims to have collected 'up to 5,000 data points on over 230 million American voters'.[9] The company attempts to identify people's personality traits to predict what kind of message is most likely to persuade people.[10]

With online microtargeting, political communications can be targeted at individuals or niche audiences, and the messages can be adapted to the recipients. A company gives an example of the possibilities for targeting niche audiences: 'targeting fathers aged 35-44 in Texas who frequent gun enthusiast websites'.[11] The digital director of the Donald Trump campaign suggests that there is not much difference between behavioural targeting and online political microtargeting: 'It's the same shit we use in commercials, just has fancier names.'[12]

---

2 W. Gorton, 'Manipulating Citizens: How Political Campaigns' Use of Behavioral Social Science Harms Democracy', (2016) 38 *New Political Science*, no. 1, https://doi.org/10.1080/07393148.2015.1125119, pp. 61-80, p. 62.
3 I. Rubinstein, 'Voter Privacy in the Age of Big Data', (2014) *Wisconsin Law Review*, no. 5, pp. 861-936, p. 882.
4 See on behavioural targeting: J. Turow, *The Daily You: How the New Advertising Industry is Defining Your Identity and Your Worth* (2011); F. Zuiderveen Borgesius, *Improving Privacy Protection in the Area of Behavioural Targeting* (2015).
5 R. Kleis Nielsen, *Personalized Political Communication in American Campaigns* (2012).
6 C. Bennett, 'Trends in Voter Surveillance in Western Societies: Privacy Intrusions and Democratic Implications.' (2015) 13 *Surveillance & Society*, no. 3/4, pp. 370-389.
7 C. Bennett, 'Voter databases, micro-targeting, and data protection law: can political parties campaign in Europe as they do in North America?', (2016) 6 *International Data Privacy Law*, no. 4, https://doi.org/10.1093/idpl/ipw021, pp. 261-275; Rubinstein, supra note 3.
8 <https://campaigngriddirect.com> and <https://cambridgeanalytica.org> (last visited 18 December 2017).
9 Cambridge Analytica, 'About us', <https://ca-political.com/ca-advantage> (last visited 8 February 2018).
10 See generally on Cambridge Analytica: P.O. Dehaye, 'Cambridge Analytica. background research', <http://www.tinyurl.com/scl-gdoc-comment> (last visited 18 December 2017).
11 Retargeter Blog, 'The power of ad targeting for politicians' (2 February 2012), <https://retargeter.com/blog/political-advertising/the-power-of-ad-targeting-for-politicians> (last visited 18 December 2017).

### 2.2 Online political microtargeting in European countries

Online political microtargeting is not yet widely deployed in European countries. However, it appears that political parties across Europe look to the practices in the US for inspiration. As Bennett notes, 'political parties elsewhere have reportedly looked with great envy on the activities of their U.S. counterparts and longed for similar abilities to find and target potential supporters and to ensure that they vote.'[13]

Parties in the UK have been quick to emulate online political microtargeting campaigns.[14] In the 2015 UK general elections, online political microtargeting helped the Conservative Party to secure key marginal constituencies and, thus, to win the elections.[15] Three major parties in the UK, the Labour Party, the Conservative Party, and the Liberal Democrats, have invested in building voter databases, with the help of consulting services and data brokers.[16]

Some political parties in Western Europe have recruited US campaign strategists to professionalise election campaigns, in particular on the use of social media and online political microtargeting.[17] To illustrate: the UK Conservatives hired Jim Messina, campaign advisor to Barack Obama, to set up microtargeting campaigns.[18] The Dutch Green Party has hired the US-based digital strategy firm Blue State Digital.[19] To sum up, microtargeting is becoming more widely used in Europe too.

## 3. Promises

We discuss some of the main promises and threats of online political microtargeting. We roughly distinguish promises and threats for citizens, political parties, and public opinion.

### 3.1 Citizens

Microtargeting could increase political participation and therefore strengthen democracy. Media in general, and social media in particular, can mobilise citizens in times of elections. Citizens may be mobilised to cast their vote on election day, attend a political event, or discuss politics with family members, friends and others. Moreover, media can increase citizens' political knowledge, and help citizens to make more informed voting choices.[20]

---

12  <https://www.bloomberg.com/news/articles/2016-10-27/inside-the-trump-bunker-with-12-days-to-go> (last visited 18 December 2017).
13  C. Bennett, 'The politics of privacy and the privacy of politics: Parties, elections and voter surveillance in Western democracies', (2013) 18 *First Monday*, no. 8, <http://firstmonday.org/ojs/index.php/fm/article/view/4789/3730#1> (last visited 18 December 2017). See also: D. Farrell, 'Campaign Modernization and the West European Party', in K.R. Luther & F. Mueller-Rommel (eds.), *Political Parties in the New Europe* (2002), pp. 63-83; B. Rottbeck, *Der Online-Wahlkampf der Volksparteien 2009: Eine empirische Analyse* (2013), p. 50; F. Hendrickx, 'Op Facebook is de verkiezingsstrijd al ontbrand', *Volkskrant.nl*, 1 October 2016 <https://www.volkskrant.nl/binnenland/op-facebook-is-de-verkiezingsstrijd-al-ontbrand~a4387173/> (last visited 18 December 2017).
14  N. Anstead, 'Data, Democracy and Political Communication – A case study examining the use of data in the 2015 UK general election', October 2015 (unpublished; on file with the authors). M. Wallace, 'The computers that crashed. And the campaign that didn't. The story of the Tory stealth operation that outwitted Labour last month', <www.conservativehome.com/thetorydiary/2015/06/the-computers-that-crashed-and-the-campaign-that-didnt-the-story-of-the-tory-stealth-operation-that-outwitted-labour.html> (last visited 18 December 2017).
15  T. Ross, 'Secrets of the Tories' election "war room"', <www.telegraph.co.uk/news/politics/11609570/Secrets-of-the-Tories-election-war-room.html> (last visited 18 December 2017); D. Wring & S. Ward, 'Exit velocity: The media election', (2015) 68 *Parliamentary Affairs*, issue suppl_1, https://doi.org/10.1093/pa/gsv037, pp. 224-240.
16  N. Anstead, 'Was this the 'social media election'? We don't know yet', in: *UK Election Analysis 2015: Media, Voters and the Campaign*, <www.psa.ac.uk/psa/news/uk-election-analysis-2015-media-voters-and-campaign> (last visited 18 December 2017).
17  Anon., 'Spinning a win. The growing cross-border trade in campaign advice', *The Economist*, 19 March 2015; Anon., 'Jim Messina: Obamas Chef-Wahlkämpfer soll der SPD helfen', *Spiegel online*, 2 February 2015; D. Farrell et al., 'Parties and Campaign Professionals in a Digital Age. Political Consultants in the United States and Their Counterparts Overseas', (2001) 6 *The International Journal of Press/Politics*, no. 4, pp. 11-30.
18  S. Payne, 'There was one pollster who predicted a Conservative victory: Jim Messina', *The Spectator*, 9 May 2015, <http://blogs.spectator.co.uk/2015/05/there-was-one-pollster-who-predicted-a-conservative-victory-jim-messina/> (last visited 18 December 2017).
19  F. Hendrickx, 'Campagnebureau Obama helpt Jesse Klaver: we kunnen leren van Nederland', *de Volkskrant*, 17 December 2017, <https://www.volkskrant.nl/binnenland/campagnebureau-obama-helpt-jesse-klaver-we-kunnen-leren-van-nederland~a4435841/> (last visited 18 December 2017).

Yet, *targeted* online information could amplify the effects of campaigns on citizens for two reasons. First, microtargeting enables politicians to engage audiences through more relevant advertisements. As an American microtargeting company puts it: 'A positive aspect of relevant campaign ads is that the ads are more relevant to the voter receiving them: voters receive ads about issues they are most likely to care about, with easily accessed links to click-through to learn more.'[21]

Traditional forms of advertising, such as television ads, reach a mass audience. But not the entire audience might be interested in such ads. Through microtargeting, specific audiences can be connected with specific agenda points of political parties. So microtargeting could lead to more relevant information or ads for specific audiences.

To illustrate: say Alice is a 20 year-old citizen and is not interested in politics. Yet Alice regularly checks her friends' Facebook updates. On Facebook, Alice receives a political ad that informs her about the viewpoints of a political party that targets younger citizens (e.g., pro-university funding). Because the political information concerns an issue that appeals to younger citizens, Alice decides to find more information about the party and its viewpoints. Thus, targeted political advertising encourages Alice to find more information, and perhaps to vote for this party.

There is a second reason why *targeted* political information can amplify the effects of campaigns. Online political microtargeting might reach citizens who are difficult to reach through mass media such as television. A challenge within democratic societies is to reach politically uninterested voters and mobilise them to participate in politics. Such citizens often opt out of traditional media exposure, such as watching television news and reading newspapers. It has been argued that those who tune out of news may not be informed about politics.[22]

However, many of these citizens may use the internet, for instance for entertainment or social media.[23] By targeting these uninterested citizens online, a political party could reach them, expose them to political information, and influence or persuade them. Such exposure increases the likelihood that citizens cast their vote or become more interested in politics. In this way, targeted political information may help to reach those who are difficult to reach in an offline environment.

In sum, online political microtargeting has possible advantages for citizens: it can reach citizens who ignore traditional media, and it can interest people in politics through tailored messages. Microtargeting might thus increase information, interest in politics, and electoral turnout.

### 3.2 Political parties

For political parties, three of the main promises of online political microtargeting are that it can be cheap, efficient, and effective. Some forms of online microtargeting can be relatively cheap for political parties. In Europe, national political parties and candidates use various social media platforms. Today, political communication routinely integrates social media 'to target messages, to recruit volunteers and donors and to track issue engagement'.[24] Compared to television broadcasting, social media can offer cheaper means

---

20  K. Holt et al., 'Age and the effects of news media attention and social media use on political interest and participation: Do social media function as leveller?', (2013) 28 *European Journal of Communication*, no. 1, https://doi.org/10.1177/0267323112465369, pp. 19-34; P. Norris, 'Tuned out voters? Media impact on campaign learning', in T. Olsson & P. Dahlgren, *Politeia Conference*, *Young People, ICTs and Democracy: Theories, Policies* (2010). See also N. Eltantawy & J. Wiest, 'Social Media in the Egyptian Revolution: Reconsidering Resource Mobilization Theory,', (2011) 5 *International Journal of Communication*, pp. 1207-1224.
21  J. Lieberman et al., 'Yes, We Can Profile You and Our Political System is Better for It,' *CampaignGrid*, 7 February 2012. Quoted in J. Turow et al., 'Americans Roundly Reject Tailored Political Advertising', Annenberg School for Communication of the University of Pennsylvania, July 2012, <www.asc.upenn.edu/news/Turow_Tailored_Political_Advertising.pdf> (last visited 18 December 2017), p. 7.
22  A. Blekesaune et al., 'Tuning Out the World of News and Current Affairs – An Empirical Study of Europe's Disconnected Citizens', (2012) 28 *European Sociological Review*, no. 1, pp. 110-126.
23  There is ample evidence that political information reaches segments of the population of lower political interest through incidental exposure, while people are using social media for entertainment purposes. Y. Kim et al., 'Stumbling upon news on the Internet: Effects of incidental news exposure and relative entertainment use on political engagement', (2013) 29 *Computers in human behavior*, no. 6, pp. 2607-2614.
24  Bennett, supra note 13.

Frederik J. Zuiderveen Borgesius, Judith Möller, Sanne Kruikemeier, Ronan Ó Fathaigh, Kristina Irion, Tom Dobber, Balazs Bodo, Claes de Vreese to communicate to a large audience.[25] Social media thus offers an alternative for small and new parties that cannot afford expensive TV campaigns to reach potential voters.[26]

A first mover advantage could accrue to the political party first starting to use microtargeting. For instance, the UK Conservative Party focused their personalised political communications on motivating their supporters to vote in particular swing constituencies during the 2015 national elections.[27] Small political campaigns on social media are a comparatively more agile form of political advertising and allow small political parties and newcomers to focus their efforts on likely supporters. This first mover advantage, however, lasts only until political competitors start using microtargeting too.

Microtargeting helps political parties to run a more efficient campaign. Instead of showing a broad range of people the same political advertisement on Facebook, campaigns can solely focus on their actual and potential constituencies. For example, a farmers' party can save money by only targeting people who live in rural areas, while ignoring obvious metropolitan users.

### 3.3 Public opinion

Regarding public opinion, microtargeting promises to increase the diversity of political campaigns, and voters' knowledge about certain issues. First, microtargeting could make political campaigns more diverse. In representative democracies, voters select political parties that they find suitable to form the government. During the election campaign, parties explain their political programme to the electorate to generate support. From a liberal perspective on democracy, election campaigns contribute to the marketplace of ideas.[28] All parties offer their political ideas and priorities to the public who can then choose the party that best fits their political ideas, preferences, and priorities. However, a key problem for voters is that the number of parties, each with a political programme, is so large that voters are overloaded with information.[29] Hence, voters choose to, metaphorically speaking, visit only a small number of market stands in the marketplace of ideas. Voters thus make their electoral decisions with limited information.

Microtargeting can expose voters to information that is most relevant for their voting decision. Many voters have specific interests in particular policy fields, for example immigration or education. With microtargeting, political parties can target voters with information within these preferred policy fields.[30] Hence, voters can base their voting decision on the programme that convinces them the most about the issue they care about the most. This would not be possible in an exclusively mass-communicated information environment. Mass-communicated campaigns are usually limited to a small number of issues that are discussed extensively by all parties. Such niche topics are unlikely to be discussed during national mass-communication campaigns.[31] Microtargeting could thus diversify political campaigns. Even though there is a smaller audience for each issue, more issues could be discussed during political campaigns. With microtargeting, topics which are only relevant to small audiences may get a market stand in the marketplace of ideas.

A potential benefit of microtargeting on public opinion is that voters can use their limited attention to process political information more efficiently, and therefore can make better-informed decisions. Thus, voters can base their decision on which candidate made the best proposal to solve the problem that is most important to them.

In the next section we discuss some of the main threats resulting from microtargeting. Again we roughly distinguish threats for citizens, political parties, and public opinion.

---

25  Ibid.
26  European Parliamentary Research Service, 'Social media in election campaigning,' Briefing, 21 March 2014, <http://www.europarl.europa.eu/RegData/bibliotheque/briefing/2014/140709/LDM_BRI(2014)140709_REV1_EN.pdf> (last visited 18 December 2017).
27  N. Anstead, 'Data-Driven Politics in the 2015 UK Election', (2017) 22 *International Journal of Press/Politics*, no. 3, pp. 294-313.
28  M. Ferree et al., 'Four models of the public sphere in modern democracies', (2002) 31 *Theory and society*, no. 3, pp. 289-324.
29  A. Downs, 'An economic theory of political action in a democracy', (1957) 65 *Journal of Political Economy*, no. 2, pp. 135-150.
30  See: Staatscommissie Parlementair Stelsel [State Committee on the Parliamentary System in the Netherlands], 'Probleemverkenning' ['Exploration of problems'], 18 October 2017, <https://www.staatscommissieparlementairstelsel.nl/documenten/publicaties/2017/10/18/probleemverkenning-staatscommissie-parlementair-stelsel> (last visited 18 December 2017), p. 49.
31  D. Hopmann et al., 'Party media agenda-setting: How parties influence election news coverage', (2012) 18 *Party Politics*, no. 2, pp. 173-191.

## 4. Threats

### 4.1 Citizens

Three threats for citizens are: their privacy could be invaded, and they could be manipulated or ignored. First, microtargeting threatens privacy. Online political microtargeting involves gathering and combining personal data about people on a massive scale to infer sensitivities and political preferences. This data gathering threatens privacy. For instance, collecting personal information can lead to chilling effects. People who suspect that their behaviour is monitored may adapt their behaviour, trying to escape attention. If people know or suspect that their website visits are tracked, they may feel uncomfortable visiting certain websites.[32] After all, by tracking people's internet use, a company can build a database of individuals and their interests. If people expect that an extreme right-wing party will win the elections, they might hesitate to visit websites about Islam.[33]

A second privacy threat concerns data breaches. Data breaches, where hackers or others access databases with personal data, are constantly in the news. To illustrate: in 2017 a marketing company contracted by the U.S. Republican Party suffered a data breach, exposing personal data of almost 200 million US citizens. 'Apart from personal details, the data also contained citizens' suspected religious affiliations, ethnicities and political biases, such as where they stood on controversial topics like gun control, the right to abortion and stem cell research.'[34] A third privacy threat is that personal data can be used for unexpected, and sometimes harmful, new purposes. Survey evidence from the US suggests that most people do not like online political microtargeting.[35]

Apart from privacy threats, there is a threat of manipulation. Politicians could use microtargeting to manipulate voters. For instance, a party could target particular voters with tailored information that maximises, or minimises, voter engagement. A party could use social media to expose xenophobic voters to information about the high crime rates amongst immigrants. Gorton warns that microtargeting 'turns citizens into objects of manipulation and undermines the public sphere by thwarting public deliberation, aggravating political polarization, and facilitating the spread of misinformation'.[36] The targeted information does not even need to be true to maximise its impact.

Political parties could also use microtargeting to suppress voter turnout for their opponents. For example, in 2016, the Donald Trump campaign reportedly targeted African-American voters with advertisements reminding voters of Hillary Clinton's earlier remarks of calling African-American males 'super predators' to suppress African-American votes.[37] Such 'dark posts' can remain hidden for people who are not targeted.[38] After all, only the targeted people see the messages.

A political party could also misleadingly present itself as a one-issue party to each individual. A party may highlight a different issue for each voter, so each voter sees a *different* one-issue party. In this way, microtargeting could lead to a biased perception regarding the priorities of that party. Moreover, online political microtargeting could lead to a lack of transparency about the party's promises. Voters may not even know a party's views on many topics.

---

32  See generally on chilling effects: N. Richards, *Intellectual Privacy: Rethinking Civil Liberties in the Digital Age* (2014), and on chilling effects in the context of online tracking: Zuiderveen Borgesius,, supra note 4, pp. 73-78.
33  Sometimes, governments seek information about people's browsing behaviour. See for instance: J. Carrie Wong & O. Solon, 'US government demands details on all visitors to anti-Trump protest website', *The Guardian* 15 August 2017, <https://www.theguardian.com/world/2017/aug/14/donald-trump-inauguration-protest-website-search-warrant-dreamhost> (last visited 18 December 2017).
34  'Personal details of nearly 200 million US citizens exposed', *BBC News*, 19 June 2017, <www.bbc.com/news/technology-40331215> (last visited 18 December 2017). See also, about a data breach in Mexico concerning more than 90 million people: Dissent, 'Personal info of 93.4 million Mexicans exposed on Amazon', 22 April 2016, <www.databreaches.net/personal-info-of-93-4-million-mexicans-exposed-on-amazon/> (last visited 18 December 2017).
35  J. Turow et al., 'Americans Roundly Reject Tailored Political Advertising', Annenberg School for Communication of the University of Pennsylvania, July 2012, <http://web.asc.upenn.edu/news/Turow_Tailored_Political_Advertising.pdf> (last visited 18 December 2017).
36  Gorton, supra note 2.
37  J. Green & S. Issenberg, 'Inside the Trump Bunker, With Days to Go', 27 October 2016, <www.bloomberg.com/news/articles/2016-10-27/inside-the-trump-bunker-with-12-days-to-go> (last visited 18 December 2017).
38  M. Funk, 'The Secret Agenda of a Facebook Quiz', *New York Times*, 19 November 2016, <www.nytimes.com/2016/11/20/opinion/the-secret-agenda-of-a-facebook-quiz.html?_r=0> (last visited 18 December 2017). See also Staatscommissie Parlementair Stelsel, supra note 30, p. 49.

Frederik J. Zuiderveen Borgesius, Judith Möller, Sanne Kruikemeier, Ronan Ó Fathaigh, Kristina Irion, Tom Dobber, Balazs Bodo, Claes de Vreese

By way of illustration, say a politician has a profile of Alice.[39] The politician has information that suggests that Alice dislikes immigrants. The politician shows Alice personalised ads. Those ads say that the politician plans to curtail immigration. The politician has a profile of Bob that suggests that Bob has more progressive views. The ad targeted at Bob says the politician will fight the discrimination of immigrants in the job market. The ad does not mention the plan to limit immigration. Ads targeted at jobless people say that the politician will increase the amount of money people on welfare receive every month. To people whose profile suggests that they mainly care about paying less tax, the politician targets ads that say the politician will limit the maximum welfare period to one year. Hence, without technically lying, the politician could say something different to each individual.[40] In sum, microtargeting could be used to manipulate voters.

A third threat for citizens is that political parties could use microtargeting to ignore certain voter groups.[41] A political campaign might not advertise to certain people, for instance because a party does not expect them to vote, or because it expects to win anyway in a certain area. Hence, certain groups might not be informed much about an election. A political campaign could ignore citizens who are not interested in politics, are unnecessary to win a particular local constituency, or are not likely to be mobilised for a particular party. As a consequence, these citizens are not exposed to the campaign. Therefore, certain groups may be underrepresented in a democracy.

In sum, online political microtargeting brings threats for citizens: they could have their privacy invaded, be manipulated, or excluded. Even if microtargeting were not effective, the mere collection of data would still be a privacy threat.

### 4.2 Political parties

Two of the main threats for political parties are that microtargeting can be expensive, and that it gives more power to intermediaries. Professional online political microtargeting can be costly.[42] Certain types of microtargeting require political parties to develop know-how; to build and maintain voter records; to collect and analyse business intelligence; to design and manage campaigns; to use digital communication channels to target voters; and to integrate all those elements in a system that enables a minute-by-minute adjustment of campaigns.

The need to commission external expertise, buy access to personal data sets, and pay service providers can quickly exceed the resources of small and new parties. Therefore, elections might be decided based on the financial resources of political parties. As Bennett notes, microtargeting 'may very well consolidate power in the larger, and more well–financed, parties and make it more difficult for smaller parties to be nationally competitive'.[43] While there could be strategic first-mover advantage for certain political parties, their adoption could create a bandwagon effect prompting other political parties to embrace microtargeting.

Second, microtargeting could make new intermediaries more powerful. This shift in the underlying logics and infrastructures of political campaigning gave rise to a new class of intermediaries that connect political parties and the electorate.[44] In recent years, a new industry has developed that provides data-driven services. Pollsters, digital strategists, social media experts, and big data consultancies measure public opinion, build and maintain voter profiles with voters' interests and anxieties, to design and test the efficiency of personalised political messages, and deliver such messages to the screens of individual voters.

The entry of new intermediaries challenges the status quo in several ways. For instance, new gate-keeper positions and bottlenecks may arise. Some intermediaries, such as social media platforms, are in a near-monopoly position in providing certain services. This gives them unprecedented power to set prices, and dictate the terms upon political parties.

---

39  This example is taken from, and includes a sentence from, Zuiderveen Borgesius, supra note 4.
40  Turow et al. warn against 'rhetorical redlining'. Turow et al., supra note 35, p. 7.
41  D. Nickerson & T. Rogers, 'Political Campaigns and Big Data', (2014) 28 *Journal of Economic Perspectives*, no. 2, pp. 51-74, <http://pubs.aeaweb.org/doi/abs/10.1257/jep.28.2.51> (last visited 18 December 2017).
42  Ibid.
43  Bennett, supra note 13. See also Staatscommissie Parlementair Stelsel, supra note 30, p. 50.
44  Bennett observes that 'in the social networking environment, the online privacy practices of political parties are also deeply dependent upon the corporate policies and technical standards and defaults of the social media companies.' Bennett, supra note 13.

Intermediaries could also provide services to political parties at their own rate and discretion, and could even refuse to deal with political parties. Such behaviour would create new types of imbalances. To illustrate: the digital strategy firm Blue State Digital says that it will never work for a certain Dutch political party.[45]

Meanwhile, old intermediaries, such as the printed press, struggle to adjust. Shrinking (political) advertising revenues drive cost cuts that can affect some of the core roles these intermediaries previously played in the political discourse, such as fact checking, in-depth reporting, etc. In sum, threats for political parties include the costs of using microtargeting professionally, and the growing power of new intermediaries.

### 4.3 Public opinion

Online political microtargeting brings several threats for the public sphere. While microtargeting could make the communication process more effective and efficient, one piece of crucial information is not being communicated: how important an issue is to the political parties themselves.

If voters receive a lot of information about one particular issue through microtargeting, they might falsely assume that the issue is one of the central issues in a political campaign. Hence, microtargeting might lead to a biased view on the issue priorities of political parties. Such a biased view is problematic, because after the elections, politicians often form coalitions and must compromise on certain policies. Microtargeting might lead to a situation in which a voter voted for a particular candidate because of his or her stance on health care, yet once in government the health-care system moves in the opposite direction, because the issue might be less central to this party than to the coalition partner. Hence, microtargeting may influence the mandate of elected politicians. As Hillygus & Shields note: 'How does a winning candidate interpret the policy directive of the electorate if different individuals intended their vote to send different policy messages? Can politicians claim a policy mandate if citizens are voting on the basis of different policy promises?'[46]

A second threat for the public sphere is the fragmentation of the marketplace of ideas. A mass-communicated political campaign is organised around a small number of issues – for example health care, terrorism, and the economy. The majority of the electorate is aware of the stances of political parties with regard to these issues. An informed public allows political parties to engage in public debates, which can lead to deliberative processes. Voters can also become part of the deliberative process by engaging in the debates. However, if citizens lose interest in these overarching issues and focus on the issues that are more relevant to them, these public debates become less democratic and deliberative processes are cut short.

In conclusion, online political microtargeting brings serious threats for citizens, political parties, and public opinion. But the most serious dangers do not have to materialise, as discussed in the next section.

### 5. Nuancing the threats

In the European context, the threats of online political microtargeting should not be exaggerated, for several reasons. First, the stricter data privacy rules in Europe (compared to the US) will probably slow down microtargeting. Second, in countries with multi-party systems, microtargeting may make less sense than in the two-party US system. And there are big differences in budgets between European campaigns and their US counterparts. Third, the influence of political marketing on voters' opinions has limits. Even if people were exposed to manipulative microtargeting, people would still learn about more general political news from other sources. We discuss each point below.

### 5.1 Legal system

In Europe, online political microtargeting might not happen on a scale like in the US, because of Europe's stricter data privacy rules.[47] Bennett suggests that the fact that microtargeting happens so much in the US can be partly explained by the absence of a general data protection law in the US.[48] For instance, data brokerage is a large industry in the US. Data brokers are 'companies that collect consumers' personal information and resell or share that information with others.'[49] In Europe, it is harder to lawfully obtain personal data from data brokers.

---

45 Hendrickx, 'supra note 19.
46 D. Hillygus & T. Shields, *The Persuadable Voter: Wedge Issues in Presidential Campaigns* (2008) p. 14.
47 See in more detail about the interplay between European data protection law and political microtargeting: Bennett, supra note 7.

Data protection law is a legal tool that aims to ensure that personal data processing only happens fairly and transparently. Data protection law imposes obligations on organisations that process personal data (data controllers), and grants rights to people whose data are being processed (data subjects). For instance, people have the right to receive information about which personal data an organisation holds about them.[50] Independent data protection authorities oversee compliance with the rules.[51]

While data protection law in Europe is generally strict, the rules for personal data about political opinions are stricter. In principle, European data protection law prohibits using personal data about political opinions, because data about people's political opinions are included in a list of 'special categories of data'.[52] There are exceptions to the prohibition. For instance, such special data may be processed when the individual concerned gives his or her explicit consent. And data protection law allows political parties, under certain conditions, to process personal data about political opinions without the individual's consent.[53] It seems plausible that data protection law in Europe makes gathering voter data more difficult than in the US. And data subjects could use their rights to obtain information about microtargeting. For example, a US academic uses his right to access his personal data to obtain information from the UK-based company Cambridge Analytica.[54]

Moreover, EU data privacy law requires transparency about personal data processing, and about most forms of online targeted marketing. Every organisation that uses personal data must offer transparency about its personal data use.[55] The organisation must disclose, for instance, the purpose of personal data processing.[56] Hence, political parties and intermediaries that collect and use personal data for online microtargeting must disclose, for instance in a privacy statement on a website, which data they use and for which purposes.

Apart from that, the EU e-Privacy Directive requires transparency and consent for the use of tracking cookies and similar files.[57] Hence, if a company uses cookies to present targeted political marketing to people, the company must inform people about the purpose of the cookie, in this case: tracking people around the internet and showing them targeted political marketing. Such transparency requirements could help to make microtargeting less opaque.

We do not claim that European data privacy law solves all privacy or election problems, or that it actually ensures that personal data are only used fairly. Compliance with, and enforcement of, data protection law are often lacking. Moreover, the law's transparency and consent requirements largely fail to inform people.[58] Additionally, the data protection rules for political parties that gather and process personal data are somewhat vague, and different member states interpret them differently.[59] Nevertheless, it seems plausible that data protection law in Europe makes microtargeting more difficult than in the US.

---

48  Ibid.
49  Federal Trade Commission, 'Data Brokers. A Call for Transparency and Accountability' (May 2014), <www.ftc.gov/system/files/documents/reports/data-brokers-call-transparency-accountability-report-federal-trade-commission-may-2014/140527databrokerreport.pdf> (last visited 18 December 2017), p. 1. See also A. Rieke et al., 'Data brokers in an open society' (November 2016), <https://www.opensocietyfoundations.org/sites/default/files/data-brokers-in-an-open-society-20161121.pdf> (last visited 18 December 2017).
50  Art. 15 of the European Parliament and Council Regulation (EU) 2016/679 of 27 April 2016 on the protection of natural persons with regard to the processing of personal data and on the free movement of such data, and repealing Directive 95/46/EC (General Data Protection Regulation), OJ L 119, 5.4.2016, p. 1. The GDPR replaces the Data Protection Directive 95/46 (with a right to access in Art. 12).
51  The sentences introducing data protection law are taken from: Zuiderveen Borgesius, supra note 4.
52  Art. 8(1) of the Data Protection Directive; Art. 9 of the General Data Protection Regulation.
53  Art. 8(2)(d) of the Data Protection Directive; Art. 9(2)(d) of the General Data Protection Regulation.
54  D. Caroll, 'Hacking the Voters. Why a British company has America's voter data and how British law can help us get it back', 1 October 2017, <https://medium.com/@profcarroll/takebackourvoterdata-21768a756672> (last visited 18 December 2017). See also: P.O. Dehaye, 'Complaint to France's Data Protection Authority against Marine Le Pen's electoral operations', 18 September 2017, <https://medium.com/@pdehaye/complaint-to-frances-data-protection-authority-against-marine-le-pen-s-electoral-operations-7e76028eb046> (last visited 18 December 2017).
55  The legal definition of 'data controller' is more complicated than 'an organisation that uses personal data': see Art. 2(d) of the Data Protection Directive; Art. 4(7) of the General Data Protection Regulation.
56  Arts. 10 and 11 of the Data Protection Directive; Arts. 13 and 14 of the General Data Protection Regulation.
57  Art. 5.3 of the e-Privacy Directive. See also Art. 8 of the proposed ePrivacy Regulation (Proposal for a Regulation of the European Parliament and of the Council, concerning the respect for private life and the protection of personal data in electronic communications and repealing Directive 2002/58/EC (Regulation on Privacy and Electronic Communications), COM(2017) 10 final, <http://eur-lex.europa.eu/legal-content/EN/ALL/?uri=CELEX:52017PC0010)>. See on that proposal: F. Zuiderveen Borgesius (ed.), 'An Assessment of the Commission's Proposal on Privacy and Electronic Communications' (June 7, 2017), study for Directorate-General for Internal Policies, Policy Department C: Citizen's Rights and Constitutional Affairs, <https://ssrn.com/abstract=2982290> (last visited 18 December 2017).

### 5.2 Electoral and political systems

There is a second reason to think that online political microtargeting will not become as big in Europe as in the US: the different electoral systems. Continental Europe's electoral systems differ from those in the UK and US. The majoritarian electoral systems in the US and the UK are characterised by a 'winner takes all' principle that makes some votes more valuable than others. In the UK majoritarian electoral system, voters directly vote for candidates in their districts. The candidate that receives the majority of votes wins the district. The elected representatives then represent their political view in parliament. In the US, every state has a number of electors. And in every state (except Nebraska and Maine), the candidate with the most votes wins all the electors.[60] Due to this system, a candidate might win the majority of the votes, but not the elections.

A majoritarian electoral system, like in the US, can lead to a situation in which specific states can decide the entire election. The outcome in many states often appears to be settled before the elections, due to the demographic set-up and preferences of the majority of the population. Other states have the potential to 'swing' from one side to the other. Political parties may only invest in convincing voters in such swing states, because those votes are most important. In most European countries, which use proportional electoral systems, the weight of the votes is spread more equally.[61] In such systems, parties have more reason to spread their funding evenly across the electorate.

In addition, political campaigns in Europe have much lower budgets than those in the US. In the US, the Hillary Clinton campaign raised over $623 million for the 2016 elections. Donald Trump's campaign raised over $334 million. Both campaigns also had funding from their parties and their 'super PACs' (political action committees), totalling their budgets respectively at $1.4 billion and $957 million.[62] By contrast, political parties in Europe typically have lower budgets. To illustrate, none of the largest political parties in Germany has a budget of more than €47 million.[63] German parties are relatively well funded in comparison with smaller European countries (the campaign budget of the largest Dutch party is less than €4 million).[64] In sum, smaller budgets probably form a barrier to invest in microtargeting.

### 5.3 Voters do not live in digital bubbles

A third reason not to overstate the threats of online political microtargeting is that people do not live in digital bubbles. Even if people were exposed to manipulative microtargeting, they would still learn about more general political news from other sources. Voters use not only advertising to learn about politics and the election campaign, but also many other sources. A recent study found that 91% of the US population had heard about the elections in a previous week. Only 1% of that group hear about the elections through a candidate app, email, or campaign website.[65] Cable TV news is still the most important information source on the political campaign in the US.[66] Hence, the electorate may still have sufficient access to non-targeted information sources that can mitigate the 'filter bubble'-related effects of online political microtargeting.[67] People also learn about a politician's views from watching TV news and from conversations with friends or colleagues. These other sources may still sufficiently inform voters about the campaign in general: the issues central to the campaign, and the priority candidates give to specific problems. More generally, it cannot be ruled out that companies that offer microtargeting services to politicians exaggerate how effective microtargeting is. In conclusion, online political microtargeting brings serious threats. On the other hand, the threats should not be exaggerated. Table 1 provides an overview of the promises and threats of microtargeting.

---

58  A. Acquisti & J. Grossklags, 'What Can Behavioral Economics Teach Us About Privacy?', in A. Acquisti et al. (eds.), *Digital Privacy: Theory, Technologies and Practices* (2007); F. Zuiderveen Borgesius, 'Behavioural Sciences and the Regulation of Privacy on the Internet', in A. Alemanno & A.L. Sibony, *Nudge and the Law – A European Perspective* (2015).
59  Bennett, supra note 7.
60  R. Turner, '"The contemporary presidency": do Nebraska and Maine have the right idea? The political and partisan implications of the district system', (2005) 35 *Presidential Studies Quarterly*, no. 1, pp. 116-137.
61  S. Birch, 'Electoral systems and party systems in Europe East and West', (2001) 2 *Perspectives on European Politics and Society*, no. 3, pp. 355-377.
62  A. Narayanswamy et al., 'How much money is behind each campaign?', *The Washington Post*, 1 February 2017, <https://www.washingtonpost.com/graphics/politics/2016-election/campaign-finance/> (last visited 18 December 2017).
63  That figure concerns 2013; the exact budgets for 2017 are not yet available. Deutscher Bundestag, 'Bekanntmachung von Rechenschaftsberichten politischer Parteien für das Kalenderjahr 2013 (1. Teil – Bundestagsparteien)', 11 March 2015, <http://dip21.bundestag.de/dip21/btd/18/043/1804300.pdf> (last visited 18 December 2017).
64  That figure concerns 2012; the exact budgets for 2017 are not yet available. VVD, 'Jaarrapport 2012', <https://vvd.nl/content/uploads/2016/12/jaarrapport2012.pdf> (last visited 18 December 2017).

*Table 1  Promises and threats of microtargeting for citizens, parties and public opinion*

|  | **Promises** | **Threats** |
| --- | --- | --- |
| **Citizens** | More relevant political advertising<br>Reaching social groups that are difficult to contact | Invading privacy<br>Manipulating voters<br>Excluding voter groups |
| **Political parties** | Cheap (some types of microtargeting)<br>Efficient<br>Effective | Expensive (some types of microtargeting)<br>More power for commercial intermediaries |
| **Public opinion** | Campaign diversification<br>More knowledge among voters about individually relevant issues | Lack of transparency regarding politicians' priorities<br>Fragmentation of the market place of ideas |

## 6. Regulating online political microtargeting

In this section, we will highlight the possibilities for national legislators which want to regulate online political microtargeting. But first we will discuss the right to freedom of expression of political parties, which may limit the possibilities for regulating microtargeting.

### 6.1 Freedom of expression and regulating online microtargeting

Any law that restricts political communication must comply with Article 10 of the European Convention on Human Rights, which guarantees the right to freedom of expression. While the right to freedom of expression is not absolute, governments may only restrict this right in very limited circumstances.[68] The European Court of Human Rights in Strasbourg, which is tasked with interpreting the Convention, ultimately decides whether government restrictions on the freedom of expression are compatible with its guarantees. Is online political microtargeting protected by Article 10?

The Strasbourg Court has held that 'publishing information with a view to influencing' voters is an exercise of freedom of expression.[69] On the basis of this principle, online political microtargeting would seem to come within the definition of freedom of expression. Indeed, a closely related means of political communication, paid political advertising on television, has also been held to be covered by Article 10.[70] Similarly, distributing election leaflets,[71] and displaying political posters,[72] are covered by the right to freedom of expression under

---

65  J. Gottfried et al., 'The 2016 presidential campaign – a news event that's hard to miss', *Pew*, 4 February 2016, <www.journalism.org/2016/02/04/the-2016-presidential-campaign-a-news-event-thats-hard-to-miss/> (last visited 18 December 2017).
66  Ibid.
67  See F. Zuiderveen Borgesius et al., 'Should We Worry about Filter Bubbles?', (2016) 5 *Internet Policy Review*, no. 1, pp. 1-16.
68  These circumstances include, for example, 'in the interests of national security', the 'prevention of disorder or crime', or the 'protection of the reputation or rights of others' (see Art. 10(2) Convention for the Protection of Human Rights and Fundamental Freedoms 213 UNTS 222).
69  *Bowman v UK*, Application No. 24839/94, Merits and Just Satisfaction, 19 February 1988, para. 47.
70  *Animal Defenders International v UK*, Application No. 48876/08, Merits and Just Satisfaction, 22 April 2013, para. 117; and *TV Vest As & Rogaland Pensjonistparti v Norway*, Application No. 21132/05, Merits and Just Satisfaction, 11 December 2008, para. 78. For a discussion of the regulatory environment for the media during elections, see M. Cappello (ed.), *Media coverage of elections: the legal framework in Europe* (2017).
71  *Andrushko v Russia,* Application No. 4260/04, Merits and Just Satisfaction, 14 October 2010, para. 42.
72  *Kandzhov v Bulgaria,* Application No. 68294/01, Merits and Just Satisfaction, 6 November 2008, para. 70.

Article 10. The Court has also held that Article 10 not only protects the 'content of information', but also the 'means of transmission or reception'.[73]

Online political microtargeting, similar to other political advertising, could be considered 'political speech', which enjoys the highest level of protection under Article 10.[74] The Court has held on numerous occasions that any restrictions on political speech are subject to the Court's most demanding standard of scrutiny, namely 'strict scrutiny'.[75] Therefore, there is usually 'little scope' for restrictions on political speech under Article 10, and any such restriction must be 'narrowly interpreted', and its necessity 'convincingly established' by the government.[76] Further, the Court has held that a political party's freedom of expression is protected by Article 10, given the 'essential role' of political parties in 'the proper functioning of democracy'.[77]

A Norwegian political party argued that a ban on political advertising on television, during the run-up to elections, violated its right to freedom of expression. In *TV Vest v Norway,* the Strasbourg Court found a violation of Article 10.[78] The Court recognised that there could be relevant reasons for a ban on political advertising, such as preventing the 'financially powerful' from obtaining an 'undesirable advantage' in public debates, and 'ensuring a level playing field in elections'.[79] However, the Court held that the political party at issue, a small pensioners' party, was 'hardly mentioned' in election television coverage, and paid advertising on television became 'the only way' for it to put its message to the public.[80] Moreover, the party did not fall within the category of a party that the ban was designed to target, namely financially strong parties which might gain an 'unfair advantage'.[81] Thus, the Court held that the general 'objectives' of the ban could not justify its application to the political party, and thereby violated its right to freedom of expression under Article 10. Therefore, the Article 10 principles set out above protecting political expression, and a political party's expression, in addition to the Court's judgment in *TV Vest*, would seem to suggest that a ban on online political microtargeting would be difficult to reconcile with Article 10.

However, in 2013, the Court, sitting in a 17-judge Grand Chamber (due to the importance of the case), held in *Animal Defenders International v UK* that a ban on paid political advertising on television in the UK did not violate Article 10.[82] But unlike *TV Vest*, the case concerned an animal rights group which sought to broadcast a political advertisement outside an election period. For the first time under Article 10, the Court held that a certain type of regulation, which the Court called 'general measures', can be imposed 'consistently with the Convention', even where they 'result in individual hard cases' affecting freedom of expression.[83]

The Court laid down a new three-step test for determining whether a 'general measure' is consistent with Article 10: the Court must assess the 'legislative choices' underlying the general measure, (b) the 'quality' of the parliamentary review of the necessity of the measure, and (c) any 'risk of abuse' if a general measure is relaxed.[84]

The Court then applied its general-measures test to the ban on political advertising on television in the UK: first, the Court examined the 'legislative choices' underlying the ban, and accepted that it was necessary to prevent the 'risk of distortion' of public debate by wealthy groups having unequal access to political advertising;[85] and due to 'the immediate and powerful effect of the broadcast media'.[86] Second, with regard to the quality of parliamentary review, the Court attached 'considerable weight' to the 'extensive pre-legislative consultation', referencing a number of parliamentary bodies which had examined the ban.[87]

---

73  *Autronic AG v Switzerland,* Application No. 12726/87, Merits and Just Satisfaction, 22 May 1990, para. 47.
74  *TV Vest AS v Norway*, supra note 70, para. 66.
75  Ibid., para. 64.
76  *Vitrenko and Others v Ukraine,* Application No. 23510/02, Decision, 16 December 2008, para. 1.
77  *Refah Partisi and Others v Turkey,* Application No. 41340/98, Merits and Just Satisfaction, 13 February 2003, para. 89.
78  *TV Vest v Norway,* supra note 70.
79  Ibid., para. 70.
80  Ibid., para. 73.
81  Ibid., para. 72.
82  *Animal Defenders International v UK*, supra note 70.
83  Ibid., para. 106.
84  Ibid., para. 108.
85  Ibid., para. 117.
86  Ibid., para. 119.
87  Ibid., para. 115.

Third, as regards the risks from relaxing a general measure, the Court held that it was 'reasonable' for the government to fear that a relaxed ban (such as financial caps on political advertising expenditure) was not feasible, given the 'risk of abuse' in the form of wealthy bodies 'with agendas' being 'fronted' by social advocacy groups, leading to uncertainty and litigation.[88] Therefore, the Court held that the total ban on political TV advertising was consistent with Article 10.

Regarding restrictions on online political microtargeting, would the Court apply the general-measures analysis from *Animal Defenders*, or the principles in *TV Vest*? This is an open question, as the Court in *Animal Defenders* did not explicitly overrule *TV Vest*.

Finally, one further point to consider when discussing whether regulation may be imposed on online political microtargeting consistently with the European Convention is Article 3 of Protocol 1 to the Convention, which guarantees a right to free elections. Article 3 imposes an obligation on governments to ensure that elections take place under conditions which 'ensure the free expression of the opinion of the people in the choice of the legislature'.[89] The Court has held that Article 10 of the Convention (freedom of expression) and Article 3 of Protocol 1 may 'come into conflict', and 'in the period preceding or during an election' it may be 'necessary' to impose 'certain restrictions, of a type which would not usually be acceptable, on freedom of expression, in order to secure the 'free expression of the opinion of the people in the choice of the legislature'.[90]

Indeed, the Grand Chamber in *Animal Defenders* recognised this principle that 'statutory control of the public debate' was necessary to 'protect the electoral process' (even though the ban applied not only during election periods, but outside these periods also). The Court added that 'the risk to pluralist public debates, elections and the democratic process would evidently be more acute during an electoral period'.[91] In sum, rules on online political microtargeting must respect the right to freedom of expression.

### 6.2 Possible regulation of online political microtargeting

It is too early to give definitive policy advice regarding online political microtargeting. We simply do not yet know enough about the practice and its effects. If research shows that the advantages clearly outweigh the disadvantages, perhaps policymakers should not do much. On the other hand, if the disadvantages outweigh the advantages, solutions must be found. Unfortunately, at the moment, the precise workings and effects of online political microtargeting are unclear.

What could be done in the short term? As a starting point for a discussion, we give some tentative suggestions. First, there is a need for more information on microtargeting, and thus for more research. For example, it is still an open question what the effects of political microtargeting on citizens are. This research should also include a normative component. We need to determine at which point we decide that the benefits of microtargeting outweigh the risks.

Second, policymakers should consider requiring more transparency regarding microtargeting from political parties. For instance, parties could be required to disclose how much money they spend on online political microtargeting. There are precedents for such rules. To illustrate, the UK electoral commission's political party expenditure database contains the spending by each party on advertising through Facebook, Twitter, and Google.[92]

Perhaps the law could require political parties to provide a copy of each online ad to a central repository. In that way, researchers, journalists, and others can see what a party promises to different people.[93] The law could also require that each political ad includes information about who paid for it, and about whether that ad is targeted or not.[94] To illustrate: in the UK, the law requires that election material identifies its promotor.[95] Transparency requirements could help to obtain more information about the extent of the problem of microtargeting. Under current law, Data Protection Authorities could use their investigative powers to examine the extent of microtargeting and, for instance, to inspect political parties' processing operations. The French Data Protection Authority has investigated personal data use by political parties,[96] and the UK Authority is investigating microtargeting practices.[97]

---

88  Ibid., para. 122.
89  Art. 3 Protocol to the Convention for the Protection of Human Rights and Fundamental Freedoms 1952, ETS No. 009.
90  *Bowman v UK*, supra note 69, para. 43.
91  *Animal Defenders International v UK*, supra note 70, para. 111.
92  See UK Electoral Commission, <http://search.electoralcommission.org.uk/Search/Spending> (last visited 18 December 2017).
93  See for such a proposal: S. Barocas, 'The price of precision: voter microtargeting and its potential harms to the democratic process', in *PLEAD '12 Proceedings of the First Edition Workshop on Politics, Elections and Data* (2012), pp. 31-36.

As a result of the investigation of the influence of the services offered by social media companies on the US presidential election campaign,[98] companies offered self-regulatory measures, for example disclosing who paid for political ads.[99] However, it is difficult for regulators, journalists, and others to monitor whether self-regulation is effective and comprehensive.

If research or experience shows that microtargeting is indeed a problem that needs a solution, more substantive regulation could be considered. For example, campaign expenditure restrictions could be imposed on political parties, placing caps on online political microtargeting. However, the European Court has found that limits on campaign expenditure may violate Article 10 in certain circumstances.[100]

Perhaps policymakers could even consider an outright ban on online political microtargeting, which could be limited to election periods.[101] Such a rule could apply only to political parties, but may then risk being circumvented through third-party online political microtargeting. This risk has been recognised by the European Court as justifying a total ban on political advertising on television.[102] A similar argument could be made about online political microtargeting. To sum up, policymakers have several options to mitigate the risks of microtargeting. However, first, we need more information about microtargeting.

### 7. Concluding thoughts

We explored the advantages and disadvantages of online political microtargeting for democracy. For citizens, such microtargeting may lead to more relevant advertising. And microtargeting could help to reach citizens who are difficult to reach through other channels. For politicians, microtargeting can be efficient, effective, and – in some cases – cheap. Microtargeting can benefit public opinion, as it can lead to more diverse political campaigns, and to more knowledge among voters about certain issues.

But microtargeting also brings disadvantages for democracy. It can invade people's privacy, and could be used to exclude or manipulate people. For example, microtargeting enables a political party to, misleadingly, present itself as a different one-issue party to different people. For politicians, microtargeting also brings threats. Some types of microtargeting are so expensive that large parties with more funding gain an unfair advantage. And new intermediaries, such as online marketing companies, become more powerful. A risk for public opinion is that the priorities of political parties may become opaque. Moreover, political discussions may become fragmented when different voter groups focus on different topics.

---

94  Similar requirements are already in place for general advertising. Several EU-wide rules require advertising to be labelled as such. See for example number 22 of the Annex of the Unfair Commercial Practices Directive 2005/29/EC; Art. 9(1)(a) of the Audiovisual Media Services Directive 2010/13/EU; Art. 6 of the E-Commerce Directive 200/31/EC. As noted in Section 5.1 of this paper, the ePrivacy Directive and the General Data Protection Regulation require transparency about most uses of personal data and targeted online advertising. The Dutch CPB Netherlands Bureau for Economic Policy Analysis also calls for more transparency: 'Platforms veranderen de wereld. Beleid voor transparantie [Platforms are changing the world. Policy for transparency]' (December 2017), <https://www.cpb.nl/sites/default/files/omnidownload/CPB-Policy-Brief-2017-11-Scientia-Potentia-Est-De-opkomst-van-de-makelaar-voor-alles.pdf >(last visited 18 December 2017).
95  Representation of the People Act 1983, s. 110.
96  Commission nationale de l'informatique et des libertés (CNIL), 'Délibération n° 2012-020 du 26 janvier 2012 portant recommandation relative à la mise en œuvre par les partis ou groupements à caractère politique, élus ou candidats à des fonctions électives de fichiers dans le cadre de leurs activités politiques', 9 February 2012, <https://www.legifrance.gouv.fr/affichTexte.do?cidTexte=JORFTEXT000025344843> (last visited 18 December 2017). See also: Bennett, supra note 7.
97  E. Denham, 'Update on ICO investigation into data analytics for political purposes', <https://iconewsblog.org.uk/2017/12/13/update-on-ico-investigation-into-data-analytics-for-political-purposes/> (last visited 18 December 2017).
98  D. Kreiss & S. Mcgregor. 'Technology Firms Shape Political Communication: The Work of Microsoft, Facebook, Twitter, and Google With Campaigns During the 2016 US Presidential Cycle', (2017) *Political Communication*, https://doi.org/10.1080/10584609.2017.1364814, pp. 1-23.
99  J. Roettgers, 'Facebook Wants to Self-Regulate Political Advertising, Provide Russian Ads to Congress', *Variety*, 21 September 2017, <http://variety.com/2017/digital/news/facebook-political-ads-1202565678/> (last visited 18 December 2017).
100 *Bowman v UK*, supra note 69.
101 The Dutch CPB Netherlands Bureau for Economic Policy Analysis mentions the possibility of prohibiting politicians from adapting messages to individual voters. CBP, 'Platforms veranderen de wereld. Beleid voor transparantie [Platforms are changing the world. Policy for transparency]' (December 2017), <https://www.cpb.nl/sites/default/files/omnidownload/CPB-Policy-Brief-2017-11-Scientia-Potentia-Est-De-opkomst-van-de-makelaar-voor-alles.pdf> (last visited 18 December 2017).
102 *Animal Defenders International v UK*, supra note 70, para. 122.

These risks are serious, and if they materialise, they threaten democracy. Nevertheless, these risks should not be overstated. Microtargeting might have less influence in Europe than in the US, because of differences in the legal and electoral systems. Moreover, the influence of online political advertising on voters has limits.

We highlighted some options for policymakers if they want to mitigate the risks. Because of freedom of expression norms, policymakers must tread carefully when regulating political speech, and when regulating political advertising. But before policy is developed, the main priority is obtaining more information about the extent and the effects of online political microtargeting. At a minimum, transparency about online political microtargeting should be improved.